\documentclass{aa}
\usepackage{graphics}

\begin{document}

\title{New constraints on the Cosmic Mid-Infrared Background
using TeV gamma-ray astronomy}

\author{C. Renault\inst{1}\thanks{Now at ISN Grenoble}, A. Barrau\inst{2}, G. Lagache\inst{3}, J.-L. Puget\inst{3}}

\offprints{C. Renault} \mail{rcecile@in2p3.fr}

\institute{
 LPNHE, CNRS-IN2P3 Universit\'es Paris VI-VII, 4 place
 Jussieu, F-75252 Paris Cedex 05, France
 \and
ISN Grenoble, CNRS-IN2P3 Universit\'e Joseph Fourier, 53 av des Martyrs, 38026
Grenoble cedex, France
 \and
IAS, Universit\'e de Paris XI,  91405 Orsay Cedex
}

\date{Received;Accepted} 

\titlerunning{New Constraints on the Cosmic MIR Background ...}

\authorrunning{Renault et al.}

\abstract{
Very high energy gamma-ray data obtained by CAT and HEGRA 
from active galactic nucleus Mkn~501 are used to 
constrain the cosmic Mid-Infrared background. 
While the entire infrared and submillimeter spectrum shape based on models has
been fixed and the density scaled as a whole in previous studies, recent measures on the low and high 
energy infrared background are extensively used in this paper. In this original approach the
infrared distribution is only varied in the unexplored 3.5-100~$\mu$m region.
With conservative hypothesis on the intrinsic 
spectrum of Mkn~501, an upper limit of 4.7~nW.m$^{-2}$.sr$^{-1}$ between 5 and 15~$\mu$m 
is derived, which is very close to the lower limit inferred from deep ISOCAM
cosmological surveys at 15 $\mu$m.
 This result is shown to be independent of the exact density of the 
\mbox{$\lambda < 3.5~\mu$m} and \mbox{$\lambda > 100~\mu$m} infrared distribution
within the uncertainties of the measurements.
Moreover, the study presented here rules out a complete
extragalactic origin for the 60 micron tentative background
detection as found by Finkbeiner et al. (\cite{finkbeiner}).                                                  
\keywords{Cosmology: diffuse radiation; Infrared: general; Gamma-rays: observations;
BL~Lacertae objects: Mkn~501}
}
   
\maketitle

\section{Introduction} 
Our knowledge of the early epochs of galaxies has recently
increased thanks to the observational evidences provided by 
UV/Vis/Near-IR, far-IR and submillimeter (submm) surveys of high-redshift objects. 
In a consistent picture, galaxy formation and evolution
can also be constrained by the background radiation which 
is produced by the line-of-sight
accumulation of all extragalactic sources. \\

In the last two years the Extragalactic Background (EB) at visible, IR and
submm wavelengths has been finally constrained by both very deep
source counts and upper limits on the diffuse isotropic emission at shorter
wavelengths, and measurements in the submm range (see Gispert et al. \cite{gispert}
for a review).\\
The Cosmic Infrared Background (CIB) detected in the COBE data at wavelength greater than 100 $\mu$m
(Fixsen et al. \cite{fixsen}, Hauser et al. \cite{hauser}, Schlegel et al. \cite{schlegel}, Lagache et al. \cite{lagache}) contains a 
surprisingly large fraction of the background due to distant galaxies. 
The spectrum implies that the submm part of the CIB cannot be dominated 
by the emission of the galaxies which accounts for most of the CIB at 150~$\mu$m, 
and thus contains a unique information about high redshift IR galaxies.
Considering the variety of long wavelength spectra observed for these 
galaxies, Gispert et al. (\cite{gispert}) have shown that a
co-moving production rate of far-IR radiation with strong evolution at low 
redshifts but little evolution between redshifts 1 and 4 is the only solution 
allowed by the CIB.\\
The search for the UV, optical, near-IR and mid-IR EB
(from 0.2 to 2.2 $\mu$m)  obtained by summing up the contributions of galaxies
using number counts currently gives only lower limits; but
the flattening of the faint counts 
suggests that we are now close to convergence.
Nevertheless, number counts give a good determination of the background
only if the full flux of the sources is properly taken into account
(Bernstein et al. \cite{bernstein})
and if diffuse components are negligible. In fact,
studies of the diffuse COBE emission
(Gorjian at al. \cite{gorjian}; Dwek \& Arendt \cite{dwek98}; Cambr\'esy et al. \cite{cambresy}) 
suggest that a significant fraction of the EB is missed from number counts integrals
in the near-IR.\\

Due to the bright zodiacal foreground emission (see
for example Kellsall et al. \cite{kelsall}), EB
determinations from 5 to about 80 $\mu$m 
are very difficult. In this wavelength range, EB constraints
come mostly from deep IR cosmological surveys as the ones performed
with ISOCAM (e.g. Elbaz et al. \cite{elbaz}). 
A more controversial tentative CIB detection at 60 $\mu$m has also been published using DIRBE data
(Finkbeiner et al. \cite{finkbeiner}).\\

It has been suggested that observations of the TeV spectrum of extragalactic sources can be a powerful
tool to constrain the CIB spectrum (Nikishov \cite{nikishov}; Gould \& Schreder \cite{gould}; 
Stecker et al. \cite{stecker92}), especially around 10~$\mu$m, where the EB constraints
are today very weak. As a matter of fact,
TeV gamma rays propagating in the intergalactic medium 
undergo absorption through electron pair production on CIB photons.
Using sources Mkn~421 and Mkn~501, meaningful upper limits have been 
established (see Malkan \& Stecker \cite{malkan}, Funk et al. \cite{funk}, Macminn \& Primack \cite{macminn}, 
Stanev \& Franceschini \cite{stanev}, Guy et al. \cite{guy} for example).\\

In this paper, gamma-ray data are used, together 
with the best presently available direct 
CIB measurements (rather than CIB models), to improve the
constraints in the most difficult spectral region, near the
maximum of the zodiacal emission. The paper is organised as follows. We discuss in Sect. 2
the CIB measurements that are going to be used. Sect. 3
presents the TeV gamma-ray data.
The determination of the upper limit is explained and done in Sect. 4 and 5.
In Sect. 6, we compare our upper limit with lower limits derived
from deep source counts obtained with the ISOCAM instrument aboard ISO
(Elbaz et al. \cite{elbaz}, Metcalfe et al. \cite{metcalfe})
and discuss the CIB distribution shape from 10 to 100~$\mu$m. Cosmological implications are then discussed.
Finally, in section 7, the consequences on the physical parameters of Mkn501
regarding the internal absorption are briefly considered.

\section{CIB density distribution}

This study aims at deriving upper limits on the minimum of the CIB distribution lying in
the Mid-IR (5-80~$\mu$m) region. In the Far-IR region, the distribution is
constrained to account for the 100~$\mu$m
determination from Lagache et al (\cite{lagache}), which is in very good agreement with
the value of Finkbeiner et al. (\cite{finkbeiner}); data at longer wavelengths 
are not relevant
for this study as they are beyond the threshold of the hardest TeV photons.
In the Near-IR region, three different hypothesis, parametrised in this work by polynomial fits
between 0.2 and 3.5~$\mu$m, are considered: the first one (dashed-dotted line in
Fig.~\ref{ir_modele}, referred as $HDF$) is directly based on Hubble Deep Field galaxy counts 
(Pozzetti~et~al \cite{pozzetti}), and can therefore be considered as a lower limit; the second 
one (full line in Fig.~\ref{ir_modele}, referred as $HDF+spectro$) is derived from 
HDF results combined with ground-based spectrometry (Bernstein~et~al. \cite{bernstein}) and the
third one (dotted line in Fig.~\ref{ir_modele}, referred as $Cambresy$) is based on
(Cambr\'esy et al. \cite{cambresy}) points at 1.25 and 2.2~$\mu$m linked to the UV point at 
0.2~$\mu$m (Bowyer et al. (\cite{bowyer}) combined with Armand et al. (\cite{armand})), as it is clearly 
incompatible with $HDF$ and $HDF+spectro$ hypothesis{\footnote{However, we will see 
(Sect. 5.2) that the impact of the density below 1~$\mu$m is $\approx$null}.
All these parametrisations are required to agree with the 3.5~$\mu$m measurement from
Gorjian et al. (\cite{gorjian}); $HDF$ and $HDF+spectro$ hypothesis also include the 1.25 and 2.2~$\mu$m points
from Wright (\cite{wright}).

The CIB coming from the line of sight accumulation of all extragalactic sources is expected 
to have a smooth spectrum. The CIB might present a minimum between the stellar dominated part
and the interstellar dust dominated one between 3 and 15 $\mu$m.
To make optimum use of the upper limits that TeV gamma rays can provide
we thus use an empirical model for the CIB with a small number
of free parameters ($\lambda_{min}$
and $f_{min}$) which interpolates between 3.5 and 100 $\mu$m
 two segments of parabola (in log-log plot) linking the 3.5 or 100 $\mu$m points
to the minimum located at  ($\lambda_{min}$, $f_{min}$) with a null first derivative at this point.
The CIB empirical density distribution is shown 
in Fig.~\ref{ir_modele} for $\lambda_{min}=$~10~$\mu$m and 
$f_{min}$= 2, 4 or 6~nW~m$^{-2}$~sr$^{-1}$.

\begin{figure}
\resizebox{\hsize}{!}{\includegraphics{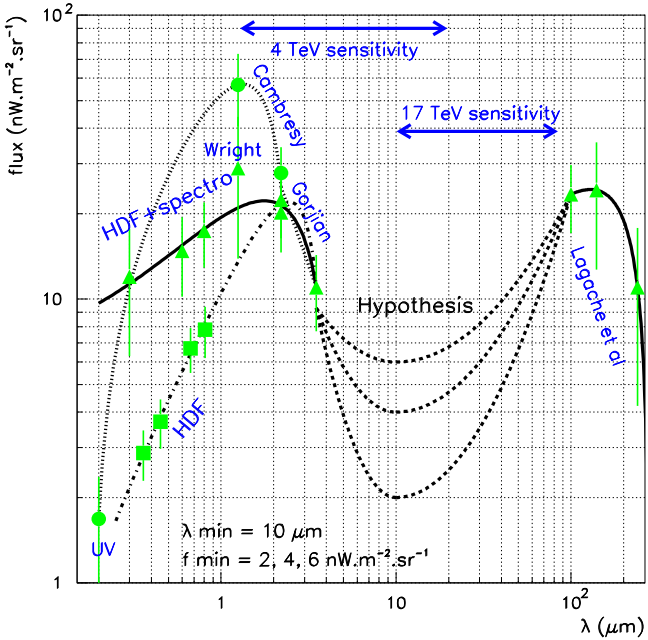}}
\caption{Assumed shapes of the CIB density distribution between 0.2 and 100~$\mu$m.
Each profile between 3.5 and 100~$\mu$m is characterised by the position of the
minimum ($\lambda_{min}, f_{min})$. In this example $\lambda_{min}$=10~$\mu$m.
See the text for details. Arrows show the 90~$\%$ interacting interval of target radiations for a 
4 or 17~TeV photon (for $f_{min}$=~4~nW~m$^{-2}$~sr$^{-1}$).
The full line links together the (HDF+spectro) points (0.3 to 0.8~$\mu$m), the (Wright  \cite{wright}) points (1.25 to 2.2~$\mu$m)
and the (Gorjian et al. \cite{gorjian}) points (2.2 to 3.5~$\mu$m). 
The dashed-dotted line links together smoothly the HDF points to the  (Wright \cite{wright}) and  (Gorjian et al. \cite{gorjian}) points.
The dotted line links together the UV measurement at 0.2~$\mu$m,
the (Cambr\'esy et al. \cite{cambresy}) points (1.25 to 2.2~$\mu$m) and the 
3.5 $\mu$m (Gorjian et al. \cite{gorjian}) points. }
\label{ir_modele}
\end{figure}

\section{TeV gamma-ray astronomy data}

The TeV source  Mkn~501 is the  second  closest X-ray selected BL~Lac 
object  after Mkn~421 with a redshift \mbox{$z \simeq 0.034$.}
During the 1997 outburst which lasted  several months,   Mkn~501  was observed intensively
in X-rays (BeppoSAX:  Pian ~et~al. \cite{pian},  RXTE: Lamer \& Wagner \cite{lamer})    and  TeV 
$\gamma$-rays
(Whipple: Catanese ~et~al. \cite{catanese},   Samuelson ~et~al. \cite{samuelson};  
HEGRA: Aharonian ~et~al. \cite{aharon97}, \cite{aharon99a},\cite{aharon99b};
Telescope Array: Hayashida ~et~al. \cite{hayashida};  
CAT:  Djannati-Ata\"{\i} ~et~al.  \cite{djannati}).  
The exceptional April~16,~1997 flare  was observed 
by  BeppoSAX and low-energy threshold ($\sim$~300~GeV)  
Whipple and   CAT atmospheric   Cherenkov telescopes,
allowing the derivation of the energy spectrum  with good accuracy in a broad dynamical
range.

These unique data  initiated  interesting efforts to set meaningful   upper limits on the 
CIB flux (see   Biller ~et~al. \cite{biller},  Stanev \& Franceschini \cite{stanev}, Barrau \cite{barrau}, 
 Stecker \& De Jager \cite{stecker98},  Stecker \cite{stecker99},
Aharonian ~et~al. \cite{aharon99b},  Coppi \& Aharonian \cite{coppi}, Konopelko ~et~al. \cite{konopelko}, Guy ~et~al. \cite{guy}). 

In this paper, CAT and HEGRA data from 400~GeV to 17~TeV are used. 
Due to bad weather, the HEGRA telescope system could not observe this  flare. 
But the
HEGRA observations of Mkn~501 revealed 
that, despite flux variations in sub-day time-scales, the 
shape of the energy spectrum above 1~TeV remained essentially stable  
throughout the entire state of high activity in 1997. The
HEGRA `time-averaged' spectrum (Aharonian ~et~al. \cite{aharon99b}) can 
therefore be used after normalisation to the
CAT April~16,~1997 spectrum at $E=$1~TeV  with the re-scaling factor of $\approx$2.2 (Guy et al. \cite{guy}). 
At these energies the 
agreement between the CAT and HEGRA spectra is quite impressive. 
Below 1~TeV, CAT points are used solely since in this energy region 
both the statistical and systematic errors of the data 
obtained close to the energy threshold of the  HEGRA 
telescope system are significantly larger than CAT data uncertainties.

\section{Unfolding TeV spectra}

The influence of low energy photons in the Universe on the propagation of 
Very High Energy (VHE) gamma-rays was pointed out by Nikishov (\cite{nikishov}). An original way of using
ground-based TeV observations of distant sources to probe the CIB was given by Stecker et al. 
(\cite{stecker92}). The fundamental idea is to look for
absorption in the intrinsic spectrum as a result of electron-positron pair production by photon
collisions $\gamma_{TeV}+\gamma_{CIB} \rightarrow e^+ + e^-$.

In such an interaction between a gamma-ray of energy
$(1+z)E$ and an infrared photon of energy $(1+z)\epsilon$, where $z$ is the redshift,
$E$ and $\epsilon$ the observed energies at $z=0$, the pair production threshold is
$E\epsilon\left(1+z\right)^2\left(1-cos\theta\right)>2\left(mc^2\right)^2$ where
$\theta$ is the angle between photons and $m$ the rest mass of the electron.
The cross-section can be written as (Heitler \cite{heitler}):
\mbox{$\sigma=k\left(1-\beta^2\right)\left(2\beta\left(\beta^2
-2\right)+\left(3-\beta^4\right)ln\left(\frac{1+\beta}{1-\beta}\right)\right) {\rm
cm}^2$}
with 
$\beta=\left(1-2\left(mc^2\right)^2/(E\epsilon\left(1-cos\theta\right))\left(1+z\right)^2\right)^{1/2}$
and $k=1.25\times10^{-25}$.
If the infrared photons have a density number $n\left(\epsilon\right)d\epsilon~{\rm
cm}^{-3}$, the corresponding optical depth for attenuation is
$$
\begin{array}{ll}
\tau(E) = & \frac{c}{H_0}\int_0^{z_s}dz\left(1+z\right)^{1/2}\\
& \int_{-1}^{1}d\left(cos\theta\right)\frac{1-cos\theta}{2}
\int_{\epsilon_t}^{\infty}d\epsilon \ n(\epsilon)\sigma(E,\epsilon,\theta)\\
\end{array}
$$
 where $\epsilon_t
=2(mc^2)^2/(E(1-cos\theta)(1+z)^2)$, $z_s$ is the redshift of the source, $c$ the
light speed, and $H_0$ the Hubble parameter (assumed here equal to 65~km~s$^{-1}$Mpc$^{-1}$). 
For this equation, the density parameter
$\Omega_0$ was chosen close to unity. The detected flux is then attenuated by a
factor $e^{-\tau(E)}$. The CIB energy distribution is assumed to be independent of $z$ as the $\gamma$-ray source redshift 
is very low (0.034).

The maximum cross-section is reached for an infrared photon
wavelength of $\lambda_{CIB}\approx \lambda_c \frac{E}{2mc^2}$ where
$\lambda_c=h/(mc)$ is the Compton wavelength of the electron. 
As shown in Fig.~\ref{ir_modele}, $\gamma$-photons with energy between a few TeV and 20~TeV ``see'' 
CIB photons with wavelengths between 3.5 and 100~$\mu$m.

\section{Upper limit on the CIB}

\subsection{Method}

To turn out the absorption corrected spectra into an upper limit on the CIB density, the intrinsic
spectral energy density of Mkn~501 is assumed to be convex in the multi-TeV region
(Guy et al. \cite{guy}). 
This conservative hypothesis is based on the fact that no natural physical process can
 re-inject energy above the Inverse-Compton bump maximum. Both Klein-Nishina effect and 
auto-absorption within the source would cause the flux to decrease more and more rapidly
as a function of energy. In the framework of more complex hadronic models this
convex shape is also expected, either because of proton-initiated-cascade (often modelized as a broken
power-law around 3 TeV), or due to the inclusion of $\mu$-synchrotron radiation (Mannheim, 
private communication). Even in the extreme case where proton synchrotron radiation is at the origin of 
the TeV bump (Aharonian \cite{aharon00}), the shape of the emission remains convex.\\

\begin{figure}
\resizebox{\hsize}{!}{\includegraphics{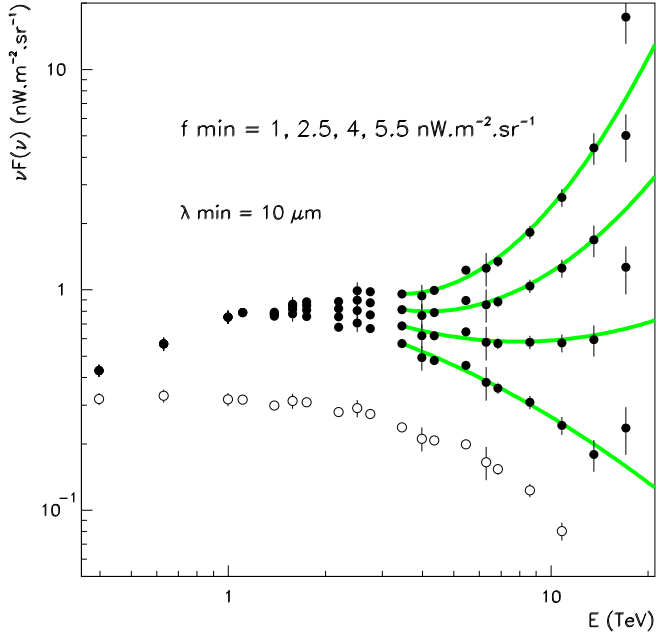}}
\caption{Shapes of the reconstructed TeV spectra.
The open circles show the observed CAT+HEGRA TeV spectrum while the filled symbols show the
absorption corrected spectra ($HDF+spectro$ hypothesis). The results of the parabola fit are superimposed.
}\label{res_nfn}
\end{figure}

Looking at the absorption corrected spectra shown in Fig.~\ref{res_nfn}, it cannot be excluded that the maximum of 
the so-called Inverse-Compton peak is not yet reached at 17~TeV when a large amount of CIB 
radiation ($f_{min}>6$~nW~m$^{-2}$~sr$^{-1}$) is assumed. Physical parameters (essentially
magnetic field and Doppler factor) required to produce such a spectral energy distribution maximum 
above 17 TeV are substantially disfavoured but, even in this case, the $\nu F(\nu)$ shape should 
remain convex. This latter point is ensured by BeppoSAX 1-100~keV
measurements (Pian et al. \cite{pian}) showing a
clearly convex spectrum before the synchrotron bump maximum 
(around 100~keV) which is supposed to be mimicked by the
TeV-spectrum before the so-called Inverse-Compton bump maximum. This hypothesis is reinforced by the fact
that the sub-TeV slope, which is independent of the CIB density beyond 3.5~$\mu$m,
effectively reflects the X-ray slope in the keV range.
This behaviour (together with the correlated variability reported {\it e.g.} by Aharonian et al. 
(\cite{aharon99a}) 
and Djannati-Ata\"{\i} et al. (\cite{djannati})) indicates that the same population of particles  is at the origin 
of both X-ray and $\gamma$-ray emissions, whatever this population is. In particular, the 
self-synchro-Compton model fits satisfactorily 
the absorption corrected data (Guy et al. \cite{guy}).\\

In order to quantify the concavity of the absorption corrected spectrum, 
a parabola fit is performed from 3.3 to 17~TeV 
in the plane ($log(\nu),log(\nu F \nu))$. 
This function, which is simple, is chosen for its constant second derivative $a= d^2(log(\nu F \nu))/d^2(log(\nu))$,
avoids the choice of a particular test-energy and fits satisfactorily the data.
 The previous physical constraint on the concavity of the TeV spectrum simply
reads as $a<0$.
The parameter $a$ is computed for $f_{min}$ in the range 0.1-7~nW~m$^{-2}$~s $^{-1}$. 
Fig.~\ref{res_nfn} shows the result of the parabolic fit in several cases, superimposed to the absorption corrected 
experimental
points. The fit under-estimates the energy density of the hardest photons,
ensuring  that the local second derivative is under-estimated and thus making the test conservative.\\

Furthermore, when $a>0$, fitted parabola with $-b/2a > 17$ TeV, where $b=d(log(\nu F
\nu))/d(log(\nu))$, are accepted. This ensures that, whatever the concavity, no 
fitted function decreasing with $\nu$ over the energy range considered
here will be excluded. Asking for a decreasing $\nu F(\nu)$ distribution is the
most natural constraint in the previously mentioned (Inverse-Compton bump maximum below
or around 1.4~TeV) most favoured case. This double test on both the first and the
second derivative makes the resulting upper limit very conservative. 

\subsection{Reliability of the upper limit}

Fig.~\ref{fig2} shows the impact of the main uncertainties on the
measured part of the CIB density distribution on the absorption corrected TeV spectrum. 
The observed spectrum of Mkn~501, combining CAT and HEGRA data, is plotted with open circles in 
panel $a)$. Spectra corrected from CIB absorption 
are superimposed with 
filled symbols in different cases: $a)$ $HDF$, $HDF+spectro$ and $Cambresy$ hypothesis
$b)$ 3.5~$\mu$m  $\pm$ 1~$\sigma$ (NIR);
$c)$ position of the minimum $\lambda_{min}=$~5 or 15~$\mu$m;
$d)$ 100~$\mu$m value $\pm$ 1~$\sigma$ (FIR).
By default, $\lambda_{min}=$~10~$\mu$m.

From panel $a)$, it can be seen that the impact of the NIR density below 1~$\mu$m is
null beyond a few TeV, as the threshold for pair production for a 1~TeV photon is 
approximately 1~$\mu$m. Furthermore, even below 1 TeV, $HDF$ and $HDF+spectro$
hypothesis lead to the same absorption corrected spectrum. Therefore, only the $HDF+spectro$ 
and $Cambresy$ hypothesis will be considered hereafter. 
The value of $\lambda_{min}$ (panel $c$)
and the FIR density at 100~$\mu$m - in a minor way - (panel $d$) can slightly change the very high energy tail of the source 
spectrum (above 10~TeV) while uncertainties on the 3.5~$\mu$m measurement 
have no sizeable impact (panel $b$). So, the uncertainty on the 100~$\mu$m measurement and a range 
of values for the parameter $\lambda_{min}$ (5-15~$\mu$m) must be taken into account
when deriving upper limits on the MIR density. 

\begin{figure}
\resizebox{\hsize}{!}{\includegraphics{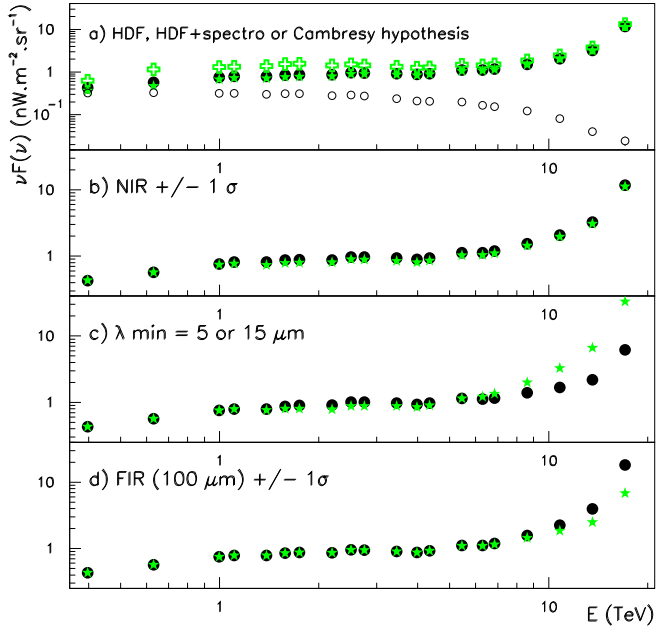}}
\caption{Impact of the shape of the CIB spectrum on the absorption corrected TeV spectrum
for f$_{min}$= 5 nW m$^{-2}$ sr$^{-1}$.
The open circles (panel $a$) show the observed TeV spectrum while the other symbols show the
absorption corrected spectra with: $a)$ $HDF$, $HDF+spectro$ and $Cambresy$ hypothesis (resp.
stars, filled circles and open crosses);
$b)$ 3.5~$\mu$m value - or + 1~$\sigma$  (resp. stars and filled circles);
$c)$ position of the minimum $\lambda_{min}=$~5 or 15~$\mu$m (resp. stars and filled circles);
$d)$ 100~$\mu$m value - or + 1~$\sigma$ (resp. stars and filled circles).
By default, $\lambda_{min}=$~10~$\mu$m. Error bars are hidden by the experimental points.
}\label{fig2}
\end{figure}

The evolution of the parameter $a$ in function of $f_{min}$ is presented in 
Fig.~\ref{para} in the case $\lambda_{min}=10\ \mu$m and a $HDF+spectro$ CIB density distribution
while the values of $f_{min}$ corresponding
to a null parameter $a$ (within 5$\sigma$) for several values of $\lambda_{min}$ are gathered in 
Tab.~\ref{tab_res} for all hypotheses regarding the NIR or the FIR density.\\

\subsection{Conclusion}

The results of this work are given in Tab.~\ref{tab_res}.
The 3.5-100~$\mu$m photons being essentially absorbed by 2-15~TeV gamma-rays, the
CAT and HEGRA spectra of Mkn~501 are particularly well suited to test the MIR density.
The high flux allows a good spectral resolution and the simultaneous X-ray data provides crucial information
on the synchrotron bump shape. The MIR upper limit derived in this paper 
is robust as it is alomost independant of the NIR  hypothesis (within 1~$\sigma$ error), and
only marginally depends on the FIR flux hypothesis (within 1~$\sigma$ error).  
Somewhat more surprising is that the exact position of the minimum in 
the 5-15~$\mu$m range significantly affects the upper limit
(as seen in Fig.~\ref{fig2}). The quite high sensitivity to the value of $\lambda_{min}$ is due to the fact that the shape 
of the $\gamma$-ray spectrum strongly depends on the CIB density distribution between 5 and 50~$\mu$m (and not only on its 
mean intensity). In other words, more absorption is allowed if the mid-IR spectral
distribution, and thus the energy dependence of the extinction, gives an acceptable
shape of the TeV spectrum of the source.

This limit is conservative, both due to the method which tends to 
under-estimate the high energy flux density and to the weak hypothesis regarding the spectrum concavity. Indeed, 
internal effects, such as Klein-Nishina cutoff and self-absorption, lead to very sharp multi-TeV intrinsic
spectra. The upper limit is more constraining than the indirect upper limit derived from the same data on Mkn~501,
which relies on a global CIB scaling (Guy et al. \cite{guy}).
Moreover, whatever the position of the minimum in the range 5-15~$\mu$m, the Mid-IR density upper limit remains
lower than 4.7~nW~m$^{-2}$~sr$^{-1}$.

Our results are in agreement with previous works: for example, Stanev \& Franceschini (\cite{stanev}) obtained an upper limit 
of 4 to 5~nW~m$^{-2}$~sr$^{-1}$ at 15~$\mu$m with two CIB hypotheses and Stecker \& De~Jager (\cite{stecker97})
derived an upper limit of 4~nW~m$^{-2}$~sr$^{-1}$ at 20~$\mu$m. It shows the "robustness" of the method
based on  TeV gamma-rays absorption by IR photons,
 essentially due to the peak in the pair production cross section. The NIR and FIR fluxes are now confirmed
by direct measurements and these observational results enhance the reliability of the constraints on the MIR intensity distribution.

An  approach similar to ours has also been proposed by Dwek et al. (\cite{dwek00}) who found a CIB density
lower than 5~nW~m$^{-2}$~sr$^{-1}$ in the range 6-30~$\mu$m and lower than 10~nW~m$^{-2}$~sr$^{-1}$
at 60~$\mu$m.

\begin{figure}
\resizebox{\hsize}{!}{\includegraphics{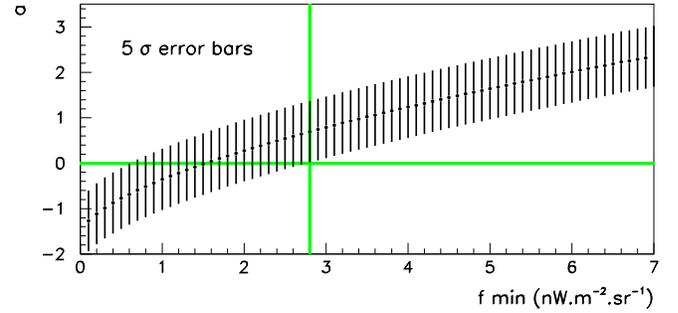}}
\caption{$a$ parameter of the parabola fit versus $f_{min}$ ($log(\nu F(\nu))=a\times
log^2(\nu)+b\times log(\nu) + c$) with 5 $\sigma$ errors (for $\lambda_{min}=10\ \mu$m
and a $HDF+spectro$ CIB density distribution.).}\label{para}
\end{figure}

\begin{table}[hhh]
\caption{5 $\sigma$ upper limit on the Mid-IR density as a function of the position of the minimum 
$\lambda_{min}$ and of the CIB density distribution.}
\begin{center}
\begin{tabular}{lllll}
\hline
& \multicolumn{3}{c}{5 $\sigma$ Mid-IR density upper limit in nW m$^{-2}$ sr$^{-1}$}  \\
IR density & $\lambda_{min}$=5~$\mu$m &  10~$\mu$m&  15~$\mu$m  \\
\hline
$HDF+spectro$ &1.3 &2.8 &4.2  \\
\hline
$Cambresy$ & 1.2& 2.6& 3.9 \\
\hline
100~$\mu$m -1~$\sigma$ & 1.8 &3.4  & 4.7  \\
\hline
100~$\mu$m +1~$\sigma$ &1.1  & 2.5 & 3.9  \\
\hline
\end{tabular}
\end{center}
\label{tab_res} \end{table}

\section{Discussion on the CIB measurements}

\subsection{The case of the 60 micron tentative EB detection}

From analysis of the DIRBE weekly averaged sky maps, Finkbeiner
et al. (\cite{finkbeiner}) have detected substantial flux in the 60 $\mu$m DIRBE channel in
excess of expected zodiacal and Galactic emission.  
The observed signal is consistent with an isotropic
background at the level $\nu I_\nu = 28.1 \pm 1.8 \pm 7(\rm{syst})$ \rm nW~m$^{-2}$~sr$^{-1}$.
While this new excess is not necessarily the CIB, they have
ruled out all known sources of emission in the solar system and
Galaxy.  They therefore tentatively interpret this signal as the CIB.
However, they point out that the IR excess exceeds limits on
the EB derived from the inferred opacity of the inter-galatic medium to observed
TeV photons. \\

In the assumption that the residual observed at 60~$\mu$m is the CIB,
we can compute, using the same method as previously,
a Mid-IR upper limit on the CIB.
With $\lambda_{min}$=10~$\mu$m, we obtain an upper limit of 1.0~nW~m$^{-2}$~sr$^{-1}$
 which is clearly incompatible with the lower limit of 3.3~nW~m$^{-2}$~sr$^{-1}$ at 15~$\mu$m,
derived from deep ISOCAM surveys (Biviano et al. \cite{biviano}). 
Even when the 60~$\mu$m tentative measurement is decreased by 1~$\sigma$,
the absorption corrected TeV spectrum in unacceptable (upper limit of 1.3~nW~m$^{-2}$~sr$^{-1}$). 
With $\lambda_{min}$=15~$\mu$m, the upper limit becomes marginally compatible with the ISOCAM result 
(upper limit of 2.2~nW~m$^{-2}$~sr$^{-1}$). 
Thus, as pointed out by Finkbeiner et al. (\cite{finkbeiner}), we conclude
that the 60~$\mu$m excess is certainly not only CIB. It contains residual
foreground emission, more likely zodiacal residual emission (see Sect. 6.3).

\subsection{The EB around 15 $\mu$m} 

Fig.~\ref{res} shows together our results with the 6.5 and 15~$\mu$m lower
limits obtained by integrating the energy from sources
observed in ISOCAM deep fields. 
The flattening of the faint counts at 15 $\mu$m (Metcalfe et al. \cite{metcalfe})
suggests that we are now close to convergence and thus that number counts 
are not very far from the true value of the EB (if no diffuse
emission contributes to it). Number counts give 3.3 $\pm$1.3 
\rm nW~m$^{-2}$~sr$^{-1}$ (Biviano et al. \cite{biviano}).
Using the model of galaxy
evolution from Dole et al. (in prep), we find an EB of about
4.4 nW~m$^{-2}$~sr$^{-1}$, which is very close to the upper
limit derived using the gamma-rays for a high value of $\lambda_{min}$. This tends to show
that the absorption in the gamma-ray spectrum
of Mkn 501 is mainly, or entirely due to absorption 
through electron pair production on CIB photons
(see Sect.~7).  Thus, the upper limit obtained with $\lambda_{min}$=15 $\mu$m 
is probably a best guest of the CIB level. 
Using a different argument based on the expected intrinsic shape of the TeV spectrum
and the ``high'' CIB model from Stecker \& De~Jager (\cite{stecker98}), Konopelko et al. (\cite{konopelko}) also concluded that 
the observed curvature of the spectrum of Mkn~501 could be naturally explained by the extragalactic absorption.

\begin{figure}
\resizebox{\hsize}{!}{\includegraphics{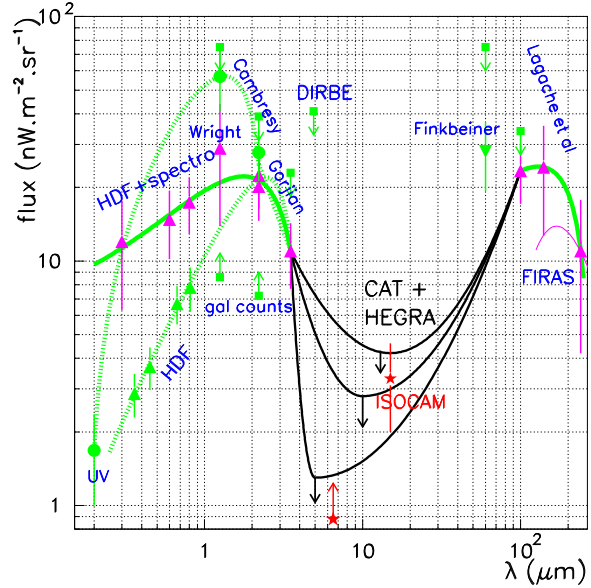}}
\caption{
CIB density in function of the wavelength: data and upper limits from VHE observation
($HDF+spectro$ case). Triangles linked by the wide full grey line
indicate the data used in this paper to 
constrain the NIR and FIR parts: below 3.5~$\mu$m
 the combination of HDF measurements and ground-based spectrometry 
(Bernstein~et~al. \cite{bernstein}), (Wright \cite{wright}) and (Gorjian~et~al. \cite{gorjian})  points and, above 100~$\mu$m,
DIRBE results (Lagache et~al. \cite{lagache}). The number counts results from ISOCAM 15~$\mu$m
(Biviano~et~al. \cite{biviano}) and 6.5~$\mu$m deep cosmological observations 
(D\'esert, private communication) are shown by stars
 while the squares come from DIRBE/COBE observations (Hauser et~al. \cite{hauser}). 
The two other hypothesis below 3.5~$\mu$m (dotted lines) lead roughly to the same upper limits.
}\label{res}
\end{figure}

\subsection{CIB shape between 10 and 100~$\mu$m} 

In the previous sections, the CIB density distribution was assumed to be described by a parabola between the minimum and 
the 100~$\mu$m Lagache~et~al. (\cite{lagache}) point. This function may underestimate of the EB light and it is important to study
the consequences of a change in the CIB shape. As illustrated in Fig.~\ref{cub}, two more realistic (and less conservative) 
distributions have been tested with a minimum at (10~$\mu$m, 2.8~nW~m$^{-2}$~sr$^{-1}$), {\it i.e.} in agreement with ISOCAM 
measurements. In each case, even taking into account only the 140~$\mu$m Lagache~et~al. point, the shape of the absorption 
corrected TeV spectrum is rejected by the physical criteria given in section 5.2. 
This means that, at least, the 100~$\mu$m Lagache et al. (\cite{lagache}) measurement may be overestimated.

\begin{figure}
\resizebox{\hsize}{!}{\includegraphics{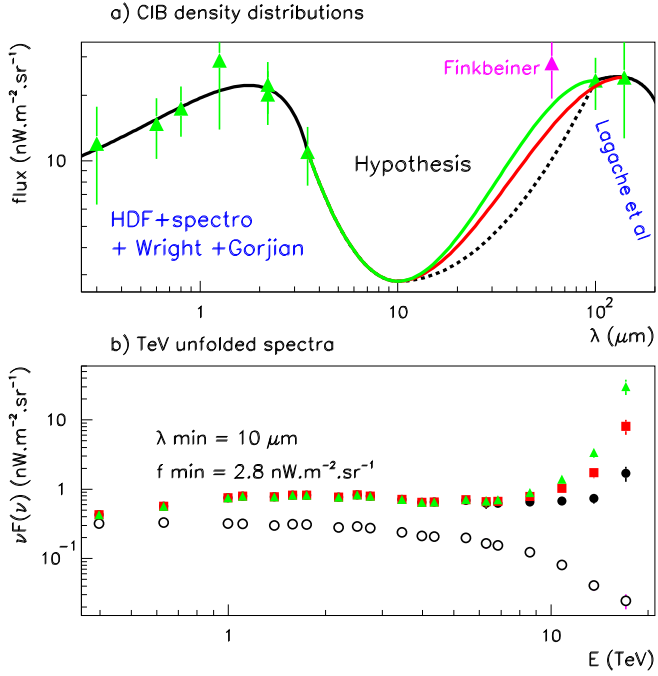}}
\caption{Consequences of a change in the CIB shape beyond~10$\mu$m: hypothesis on the CIB density distribution (panel a) and corresponding
absorption corrected high energy Mkn~501 spectra (panel b). Ordered by increasing densities, we consider the {\it normal} ({\it i.e.}
parabolic) distribution 
(dots) and two cubic distribution linking the 10~$\mu$m minimum to the 140~$\mu$m Lagache~et~al point (squares) and
the 100~$\mu$m Lagache~et~al point (up-triangles).
}\label{cub}
\end{figure}

The DIRBE zodiacal emission model was obtained by Kelsall et al. (\cite{kelsall})
relying on its time variability{\footnote{Other more empirical models have also been built
by Wright (\cite{wright}) and Finkbeiner et al. (\cite{finkbeiner}).}}. This model is critical to get the CIB in 
the near infrared. Its accuracy can be estimated using the residuals observed at wavelengths were the 
zodiacal emission is maximum (12 and 25~$\mu$m). 
The residual emission, obtained by Hauser et al. (\cite{hauser}), 
has in fact a spectrum very similar to the zodiacal one.
The residuals are about 470 nW~m$^{-2}$~sr$^{-1}$ 
at 12 and 25~$\mu$m.  The work presented just before
reveals that this best model may underestimate
the zodiacal emission as the uncertainties on other contributions (instrumental and 
interstellar) are significantly smaller. We can thus make a conservative estimate 
of the amount of zodiacal
emission which is not removed by this model at 12 and 25~$\mu$m to be about 
400~nW~m$^{-2}$~sr$^{-1}$. The amount not removed at 60 and 100~$\mu$m is thus
40~nW~m$^{-2}$~sr$^{-1}$ and 8.4~nW~m$^{-2}$~sr$^{-1}$
(using the Kelsall et al. \cite{kelsall} smooth high latitude zodiacal cloud
colour ratios). 
This reduces the CIB at 100 $\mu$m from 23.4 to 15~nW~m$^{-2}$~sr$^{-1}$. At 60 $\mu$m
the extra-zodiacal emission to be removed (40~nW~m$^{-2}$~sr$^{-1}$) 
is comparable to the residuals and thus no meaningful value can be obtained on the 
CIB at this wavelength. Using the same arguments, the extra-zodiacal emission to be removed
at 1.25~$\mu$m in the Cambr\'esy et al. (\cite{cambresy}) work
is 38.0~nW~m$^{-2}$~sr$^{-1}$ from the published result of 56.9~nW~m$^{-2}$~sr$^{-1}$.\\

The impact of a larger zodiacal emission
on the upper limits in the MIR range is shown in Fig.~\ref{res_100mu}. With a flatter CIB distribution, the upper limit 
at 10~$\mu$m increases from 2.8 to 3.7~nW~m$^{-2}$~sr$^{-1}$. It remains very close to the direct measurement
from ISOCAM at 15~$\mu$m.

\begin{figure}
\resizebox{\hsize}{!}{\includegraphics{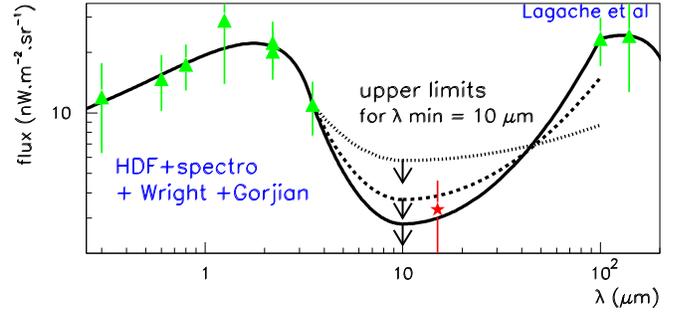}}
\caption{
CIB density $vs$ wavelength: data and upper limits from VHE observation
($\lambda_{min}$=~10~$\mu$m case). The $HDF+spectro$ spectrum is linked to different points at 100~$\mu$m:
Lagache et al. (\cite{lagache}) flux (23.4~nW~m$^{-2}$~sr$^{-1}$, full line),
the same Lagache et al. flux with a higher zodiacal background subtracted (15~nW~m$^{-2}$~sr$^{-1}$, dashed line)   and
this point~-1~$\sigma$ (8.7~nW~m$^{-2}$~sr$^{-1}$, dotted line). The number counts result from ISOCAM 15~$\mu$m
(Biviano~et~al. \cite{biviano}) is shown by the star.
}\label{res_100mu}
\end{figure}

In addition to a high interest for cosmology and galaxy evolution,
a better knowledge of the entire CIB shape would be of great interest
to unfold TeV spectra of extragalactic sources.

\section{Discussion on Mkn501 physical parameters}

In the previous section, so as to derive a conservative upper limit on
the CIB density, no internal absorption within the source has been taken into
account. Nevertheless, the ISOCAM 15~$\mu$m point (at 3.3~nW~m$^{-2}$~sr$^{-1}$) gives a lower limit on
the CIB density and, assuming this value to be a measurement, the resulting maximum possible absorption within the source can be
estimated. The previous physical constraints on the absorption corrected spectrum can be applied 
again, assuming the minimum possible CIB density (to remain conservative), and varying 
the optical depth $\tau_{AGN}$ due to $\gamma - \gamma$ pair production within the jet of Mkn501.\\

To compute $\tau_{AGN}$ in the TeV range, the low-energy spectrum ({\it i.e.} the energy
distribution of target photons) has to be known. We use the approximation derived 
by (Bednarek \& Protheroe, \cite{bednarek}) using a fit to the BeppoSAX observations made during
the April 15/16 flaring activity together with an indication from OSSE observation made
during the high state in 1997 April that the spectrum continued to approximately 500~keV
with the same energy flux per log energy interval. Performing a numerical integration of
the optical depth formula (cf Sect.~4), the authors show that, in the framework of
homogeneous SSC model, :
$$\tau_{AGN}\approx 3\times 10^8 D^{-4.8} E^{0.4} t_{var}^{-1}$$
where $D$ is the Doppler factor of the blob where relativistic electrons are assumed to 
be confined, $t_{var}$ is the variability time-scale of the source in seconds and
$E$ is the $\gamma$-rays energy in TeV. The efficiency
 usually associated with this mechanism has been supposed, as usual, equal to
unity.\\

Both absorptions are compared in Fig~\ref{selfabs}. They look rather similar as the dominant behaviour is in $E^{-2}$
in energy density. But the internal absorption is continuously increasing with energy while the CIB absorption
exhibits a ``plateau'' around 1-4 TeV.
Adding this new optical depth to the one associated with $\gamma-{CIB}$ interactions, we
derive a 5$\sigma$ upper limit for $3\times 10^8 D^{-4.8} t_{var}^{-1}$ of 0.6 (computed with $\lambda_{min}$=15 $\mu$m
and $f_{min}$=3.3~nW~m$^{-2}$~sr$^{-1}$, {\it i.e.} for the ISOCAM point). With a
reasonable variability time-scale around 1 hour, the resulting minimum Doppler factor
is $D>11.7$. Despite the large CIB absorption and the resulting stringent upper limit on the
internal optical depth, these Doppler factors are in agreement with classical AGN
descriptions and the simple homogeneous SSC model works satisfactorily. Nevertheless,
it should be pointed out that this lower limit on the Doppler factor increases as 
the wavelength of the minimum of CIB density decreases and becomes infinite below 
$\approx$ 10 $\mu m$ when there is no space left for internal absorption, unless an inhomogeneous jet is assumed.
\\
With the next generation of instruments (VERITAS and MAGIC in the northern hemisphere, HESS and CANGAROO
in the southern one) which will start operating in 2002/2003, good quality spectra of several AGNs
at different redshifts, will be available. It will lead to, simultaneously, a better understanding of the TeV emission
mechanism and a better knowledge of the CIB density distribution.

\begin{figure}
\resizebox{\hsize}{!}{\includegraphics{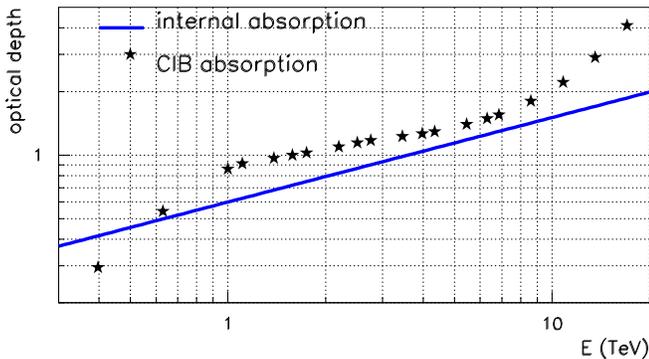}}
\caption{Optical depth of the TeV flux in function of the energy $E$.
The stars symbolise the absorption due to the EB photons while the line shows the absorption due 
to IR photons in the inner jet ($\lambda_{min}$=15~$\mu$m, f$_{min}$=3.3~nW~m$^{-2}$~sr$^{-1}$).
}\label{selfabs}
\end{figure}

\begin{acknowledgements}
The authors appreciate the contribution of the anonymous referee to the improvement of the paper.
We are grateful to  H.~Krawczynski who provided us
with  numerical values of HEGRA fluxes and with W.~Bednarek for very helpful discussions.
\end{acknowledgements}

\section{Note added in proof}

In this paper an early version of the Cambr\'esy et al. \cite{cambresy} paper was used: the 1.25~$\mu$m flux was overestimated by
5~\%; it does not affect our results.

\end{document}